\documentclass[10pt,a4paper,reqno]{amsart} 
\usepackage{psfrag}
\usepackage{epstopdf}
\usepackage{amsmath}
\usepackage{amsfonts}
\usepackage{bbm}
\usepackage{mathtools}
\usepackage{amssymb}
\usepackage{graphicx}
\usepackage{color}
\usepackage{latexsym}
\usepackage[utf8]{inputenc}
\usepackage[T1]{fontenc}
\usepackage{comment}
\usepackage{stmaryrd}
\usepackage{tikz}
\usetikzlibrary{arrows}
\renewcommand{\>}{\rangle}
\newcommand{\cs}{{\mathbb C}} 
\newcommand{\rs}{{\mathbb R}} 
\newcommand{\bexp}{b}
   
\def\e{{\rm e}}

\def\be{\begin{equation}}
\def\ee{\end{equation}}

\def\x{x_{\rm c}}
\def\y{y_{\rm c}}
\def\ys{y^*}

\newcommand{\BC}{k_{\rm B}}

\newtheorem{prop}{Proposition}

\newcommand{\E}{\mathbb{E}}

\DeclareMathOperator{\const}{const.}

\newcommand{\cT}{\mathcal T}

\DeclareMathOperator{\vv}{v}

\newtheorem{Theorem}{Theorem}
\newtheorem{Proposition}[Theorem]{Proposition}

\newcommand{\beq}{\begin{equation}}
\newcommand{\eeq}{\end{equation}}

\newcommand{\gf}{generating function}

\def\emm#1,{{\em #1}}

\graphicspath{{Figures/}}

\catcode`\@=11
\def\section{\@startsection{section}{1}%
 \z@{.7\linespacing\@plus\linespacing}{.5\linespacing}%
 {\normalfont\bfseries\scshape\centering}}

\def\subsection{\@startsection{subsection}{2}%
  \z@{.5\linespacing\@plus\linespacing}{.5\linespacing}%
  {\normalfont\bfseries\scshape}}

\def\subsubsection{\@startsection{subsubsection}{3}%
 \z@{.5\linespacing\@plus\linespacing}{-.5em}
  {\normalfont\bfseries\itshape}}
\catcode`\@=12

\def\cT{\mathcal{T}}

\begin{document}
\title
[Self-avoiding walks]
{Self-avoiding walks and polygons--an overview}

\author[A. J. Guttmann]{Anthony J. Guttmann}
\address{AJG: Department of Mathematics and Statistics, The University of Melbourne, Victoria, 3010, Australia}
\email{guttmann@unimelb.edu.au}
%


\date{\today}
\maketitle
\begin{abstract}
This is a rather personal review of the problem of self-avoiding walks and polygons. After defining the problem, and outlining what is known rigorously and what is merely conjectured, I highlight the major outstanding problems. I then give several applications in which I have been involved. These include a study of surface adsorption of polymers, counting possible paths in a telecommunication network, hitting probabilities of SAWs in a rectangle, and the modelling of biological experiments on polymers. I hope to show that SAWs are not only of intrinsic mathematical interest, but also have many interesting and useful applications.
\end{abstract}
\section{Introduction}
The problem of self-avoiding walks is one of deceptive simplicity of definition, hiding malevolent difficulty of solution. The problem was introduced by two theoretical chemists, Orr \cite{O47} and Flory \cite{F49}, as a model of a polymer in dilute solution. It soon became an interesting combinatorial model to mathematicians, and a canonical model of phase transitions, of interest to mathematical physicists. It is also a simple model of a non-Markovian process. Attempts to count the number of SAW have led to the development of new algorithms, with widespread applicability, while many more applications were discovered. These include application to the design of telephone networks, the folding and knotting of biological molecules, and a variety of chemical phenomena. Attempts at a solution have driven several mathematical advances, including developments in stochastic differential equations and probability theory.

Nearly 70 years after the model was proposed, we have a huge amount of numerical information, a substantial amount of exact information -- that is to say, results that are universally believed, but remain unproved -- and a very small body of rigorous results. In contrast, some other canonical models of phase transitions, such as the Ising model, the Potts model and percolation have either been solved (the Ising model) or much has been rigorously proved. In this short article I will outline the development of the subject, give some applications, and show that we appear to be on the verge of some major breakthroughs, which will result in proofs of much of the exact, but unproved, information that currently exists. Unfortunately all the exact and conjectural information we have applies only to the model on a two-dimensional lattice. In the case of three dimensions, we only have numerical results. Except where otherwise stated, this article will discuss the two-dimensional situation.

Finally, a veritable treasure trove of rigorous
results  would be unlocked if we could prove 
that, in the large size limit, more precisely the {\em scaling limit}, that two-dimensional random SAWs are
described by one of the SLE processes (Schramm-Loewner Evolution),
which, in the past  20 years, have been proved to describe the limit
of several discrete models in combinatorics and statistical physics.
Indeed, Lawler, Schramm and Werner proved that, if
the scaling limit of SAWs exists and is conformally invariant, then
this limit has to be $\hbox{SLE}_{8/3}$. This has been checked
via simulations by Kennedy \cite{K04}. This would in particular imply the
conjectured values of the exponents $g$ and $\nu$, which are defined below, in the two-dimensional case. We explain this in more detail in the conclusion.
\section{What is known and what isn't.}
\subsection{Self-avoiding walks}
\begin{figure}[ht!]
\begin{center} 
\includegraphics[width=4cm]{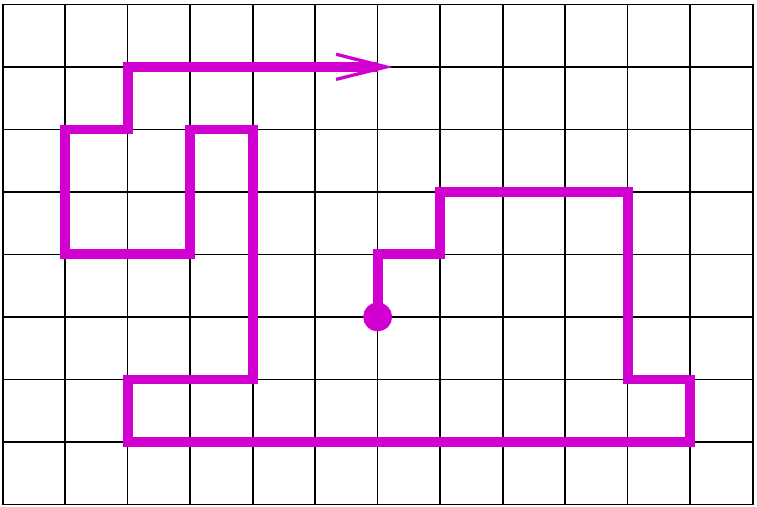}
\caption{A self-avoiding walks on the square lattice.}\label{fig:saw} 
\end{center}
\end{figure}

\begin{figure}[ht!]
\begin{center} 

\includegraphics[width=5cm]{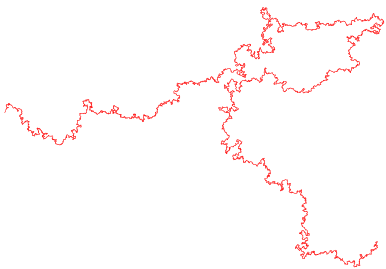}

\caption{A typical two-dimensional SAW of $2^{25}$ steps on the square lattice --
  courtesy of Nathan Clisby.}\label{fig:23D} 

\end{center}
\end{figure}

A self-avoiding walk (SAW)  of length $n$ 
on a periodic graph or lattice                      ${\mathcal L}$ 
is a sequence of distinct
vertices $w_0,w_1,
\ldots ,w_n$ in ${\mathcal L}$ 
such that each vertex is a nearest
neighbour of its predecessor. In  Figure~\ref{fig:saw} a short SAW on the square lattice is shown,
while in Figure~\ref{fig:23D} a rather long walk of $2^{25}$ steps is shown (generated by a Monte Carlo algorithm \cite{C10, C11}).

\subsubsection{How many self-avoiding walks are there?}

Two obvious questions one might ask are (i) how many SAWs are there of length $n,$ (typically defined up to translations) denoted $c_n$, and (ii) how big is a typical $n$-step SAW? Indeed, how might we measure size? A third important, but less obvious question, asks ``what is the scaling limit of SAWs?''

Frequently one rather considers the associated generating function $$C(x) = \sum_{n \ge 0} c_n x^n.$$ To see the difficulty of this problem, the reader is invited to try and calculate the first few terms $c_n$ on ${\mathbb Z}^2.$ We take $c_0$ to be 1, then $c_1=4$ as a one step walk can be in any of 4 directions. Then $c_2=12,$ $c_3=36$ and $c_4 =100.$ It is at the stage of 4-step SAWs that the self-avoiding constraint first manifests itself, and the problem becomes increasingly difficult thereafter. 

As we prove below, $c_n$ grows exponentially. Accordingly, an enormous amount of
effort has been expended 
over the last 50 years
in developing
efficient methods for counting SAW. For the square lattice, Jensen 
\cite{J12}
has extended the known series to 79 step
walks, for which he finds $c_{79} =
10194710293557466193787900071923676.$ Methods for calculating these
astonishing numbers are quite complicated (see \cite{G08}, Chapter 7), but the best current algorithm 
 still involves a counting problem of  exponential complexity,
of about 
$1.3^n$ (while a  direct counting algorithm would have complexity
$2.64^n$).

One of the few
properties one can readily prove, by virtue of the obvious
sub-multiplicative inequality 
$ c_{n+m} \le c_n c_m,$ is that the number $c_n$ grows exponentially. From this inequality
it follows that 
$$ \mu := \lim_{n \to \infty} c_n^{1/n} = \inf_n c_n^{1/n}$$
exists \cite{madras-slade}, and further that $c_n \ge \mu^n.$ 

However even the value of this ``growth constant'' $\mu$ is difficult to calculate exactly. Only in 2010
was $\mu$
 for one two-dimensional lattice, the honeycomb lattice,
actually proved by Duminil-Copin and Smirnov \cite{DCS10} to be 
$\sqrt{2+\sqrt{2}}$ (see Section~\ref{sec:DC-S}).
 For other lattices in two dimensions, and all lattices in higher dimensions, we only have numerical
estimates. For example, for the square lattice the best current
estimate is $\mu = 2.6381585303 \pm 2 \times 10^{-11},$ given by Clisby and  Jensen \cite{CJ12}. 

In fact it is believed that, for dimensionality $d > 1$ and $d\not =4$,
 $$c_n \sim \const \times   \mu^n n^g.$$
The {\em  critical exponent} $g$ is believed to 
depend on the dimension, but not on the details of the lattice. 
In particular, it is predicted to be a rational number, namely $ 11/32$,
in two dimensions. 
In three dimensions, the best estimate we have is $g = 0.156957 \pm 0.000009 $
 given by Clisby \cite{C12}.  There is no reason to believe that this number is rational.

Despite these accurate estimates, 
 we still
cannot even prove the existence of this exponent for $d < 5$, let alone establish its
value rigorously.  For $d > 4$ the higher dimensionality means that the self-avoiding restriction is less confining than in lower dimensions, and indeed has no effect on the dominant asymptotic behaviour, with the result that the SAW behaves as a random walk. More precisely, Hara and Slade \cite{HS92a, HS92b} have proved that $g=0$ in this case, and that the scaling limit is Brownian motion.
In four dimensions the above expression for $c_n$ must be modified by an additional multiplicative factor $(\log n)^{1/4},$ with $g=0.$ The appropriately rescaled walk is also expected to have Brownian motion as its scaling limit.
These assertions for the four-dimensional case are believed to be true, but no proof exists. See  \cite{BS10} for a discussion of this case.
Bounds established 50 years ago by Hammersley and Welsh \cite{HW62} have hardly been improved upon. They proved that, for SAW in dimensionality $d \ge 2,$
$$\mu^n \le c_n \le \mu^n e^{\kappa\sqrt{n}}.$$ The lower bound follows immediately from sub-additivity, while the upper bound depends on an {\em unfolding} of the walk. The number of possible unfoldings can be bounded by the number of partitions of the integer $n,$ which has the exponential behaviour given above. Note that the existence of a critical exponent would imply behaviour $\mu^n e^{\kappa\log{n}},$ which is rather far from the upper bound. A year later, Kesten \cite{K63} slightly improved the upper bound for $d > 2$ to $$c_n \le \mu^n e^{\kappa n^{2/(d+2)}\log{n}}.$$

\subsubsection{How large is a typical self-avoiding walk?}
Another important measure of SAW is the average size of 
a SAW 
of length $n$, taken uniformly at random. 
The most common measure is
the mean-square end-to-end distance, which
is believed to behave as 
$$
\E_n(|w_n|^2)
\sim \const   n^{2\nu},$$ 
(for lattices in dimensions other than 4), 
where $\nu$ is another critical
exponent. Again, its existence hasn't been proved for $d < 5$, but it is accepted
that for two-dimensional lattices $\nu = 3/4$.
 In three dimensions the best numerical estimate is $\nu = 0.587597 \pm0.000007$ \cite{C11}. In four dimensions it is believed that $\nu = 1/2,$ and again one expects a multiplicative factor $(\log n)^{1/4}.$ Finally, for $d > 4$ it has been proved \cite{HS92a} that $\nu=1/2.$ Rigorous results about $\E_n(|w_n|^2)$ are almost non-existent. It would seem intuitively obvious that
$$ cn \le \E_n(|w_n|^2) \le Cn^{2-\epsilon},$$ but only this year, in an important calculation, has substantial progress in establishing the upper bound been made by Duminil-Copin and Hammond \cite{DC-H} for SAW in dimension two or greater. They prove that walks are sub-ballistic, in the sense that there is an exponentially small probability of them having any given positive speed. But technically, the upper bound still remains an open problem\footnote{I am grateful to Alan Hammond for clarification as to precisely what has been proved.}.

\subsection{Self-avoiding polygons}
\begin{figure}[ht!]
\begin{center} 
\includegraphics[width=4cm]{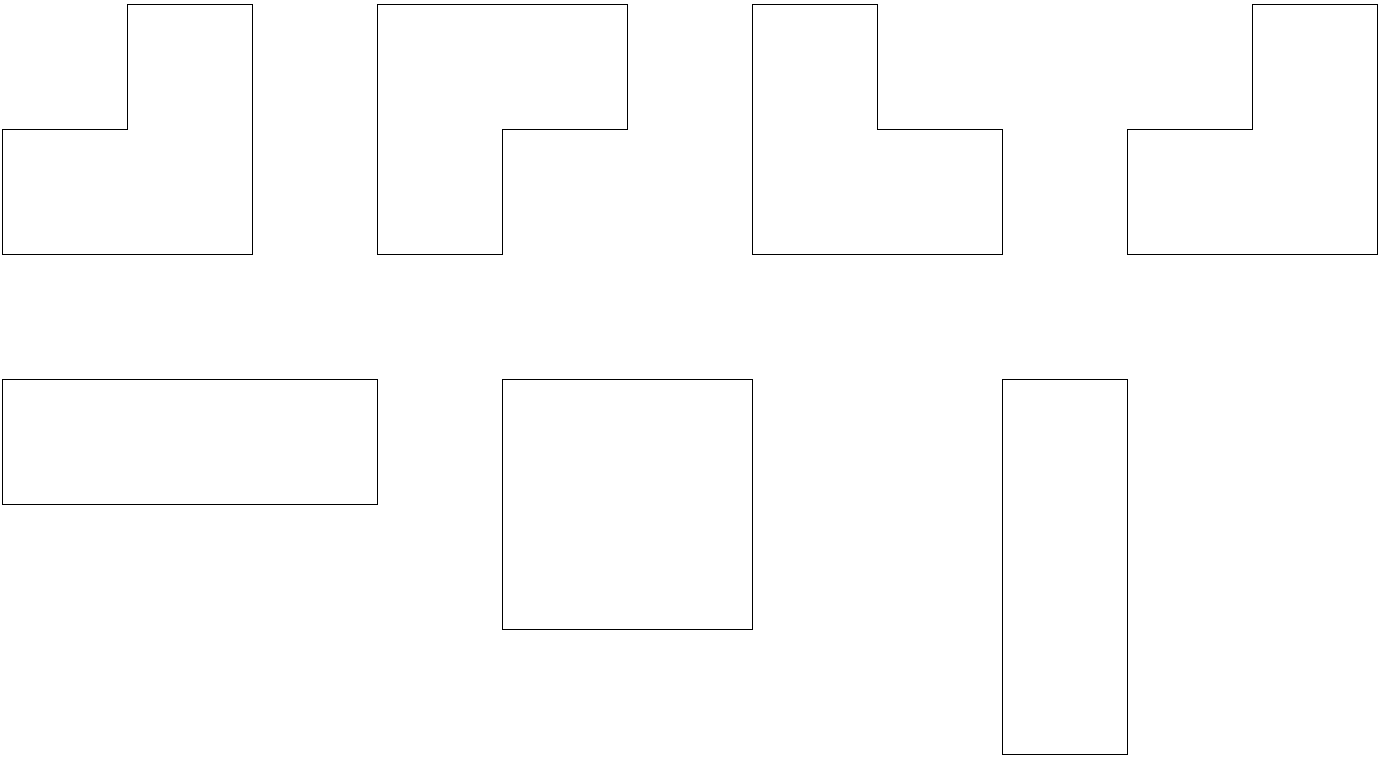}
\caption{All seven 8-sided polygons on the square lattice.}\label{fig:p8} 
\end{center}
\end{figure}

If the end-point of a SAW is adjacent to the origin, an additional step
joining the end-point to the origin will produce a self-avoiding circuit. If we ignore knowledge of the origin, 
and distinguish circuits only by their shape, we refer to {\em self-avoiding polygons} (SAP).
On the square lattice, the first non-zero embedding of a SAP is
the unit square, of perimeter 4 and area 1. There are two 6-sided polygons of area 2, and seven 8-sided polygons, shown in Figure \ref{fig:p8}, one of which has area 4, and six of which have area 3. 

Clearly, SAPs are a subset of SAWs. They are a particularly interesting subset for at least two reasons. Because the conjectured exponents for SAPs (discussed below) are integers or half-integers (which is not the case for SAWs), it is hoped that this means the underlying solution for the SAP case is simpler. Secondly, by including a second parameter, that of area, SAPs can be used to model a range of biological phenomena, such as cell inflation and collapse \cite{FGW91}.

Denote by $p_m$ the number of SAPs of
perimeter $m$, by $a_n$ the number of SAPs of area $n$, and by $p_{m,n}$ the number of SAPs of
perimeter $m$ and area $n.$ We can define two single variable generating functions, for
perimeter and area\footnote{Clearly the {\em area} of a polygon is a concept peculiar to the two-dimensional case.} respectively, and a two-variable generating function, as follows:

$$P(x) = \sum_m p_m x^m$$
$$A(q) = \sum_n a_n q^n$$
$${\mathcal P}(x,q) = \sum_{m,n} p_{m,n}x^mq^n.$$

Hammersley \cite{H61}  proved that the number of SAPs, like SAWs, grows exponentially; more
precisely 
$$\mu = \lim_{m \to \infty} p^{1/2m}_{2m}.$$
While it is far from obvious, Hammersley also proved that the growth constants $\mu$ that arise in the polygon
case and the walk case are identical. While unproved, a much stronger
result is widely believed to hold, namely that
\begin{equation}\label{1.1}
p_m \sim \const \times \mu^m m^{\alpha-3}
\end{equation}
where $\alpha$ is a {\em critical exponent}~\footnote{Note that $p_{2m+1} = 0$ for SAP on ${\mathbb Z}^d$, as only polygons
with even perimeter can exist on those lattices. For such lattices the above
asymptotic form is of course only expected to hold for even values of $m.$ For so called
close-packed lattices, such as the triangular or face-centred cubic lattices,
polygons of all perimeters greater than two are embeddable, so eqn. (\ref{1.1}) stands as
stated.}. The exponent $\alpha$ is related to the exponent $\nu$ defined above through the hyper-scaling relation $d\nu = 2-\alpha.$ This equation has not been proved, but follows from physical arguments, and of course the assumption that the exponents exist. It therefore follows from the result for $\nu$ quoted above that in three dimensions $\alpha = 0.237209 \pm 0.000021.$

For polygons there is a second growth constant, and exponent, associated with the area generating function. By concatenation arguments it can be readily proved that $$ \lambda = \lim_{n \to \infty} a_n^{1/n}$$ exists. It is also generally accepted, but not proved, that $$a_n \sim const \times \lambda^n n^\tau.$$ Unfortunately we only have numerical estimates of $\lambda$ and $\tau$ \cite{G08}. However for two-dimensional lattices $\tau$ is believed to be $-1,$ corresponding to a logarithmic singularity of the generating function. That is to say,
$$A(q) \sim const \times \log(1-\lambda q),$$ so that $a_n \sim const \times \lambda^n/n.$

Of great interest is the two-variable generating function $\mathcal{P}(x,y).$ From this, we can define the free energy
$$\kappa(q) = \lim_{m \to \infty} \frac{1}{m} \log \left( \sum_n p_{m,n} q^n \right).$$ It has been proved 
\index{free energy}
\cite{FGW91} that the free energy exists, is finite, log-convex and continuous for $0 < q < 1.$ For $q > 1$ it is infinite. The radius of convergence of ${\mathcal P}(x,q)$, which we denote $x_c(q)$, is related to the free energy by $x_c(q) = e^{-\kappa(q)}.$ This is zero for fixed $q > 1.$  A plot of $x_c(q)$ in the $x-q$ plane is shown (qualitatively) below. For $0< q < 1,$ the line $x_c(q)$ is believed to be a 
line of logarithmic singularities of the generating function ${\mathcal P}(x,q).$ The line $q = 1$, for $0 < x < x_c(1)$ is believed to be a line of {\it finite} essential singularities \cite{FGW91}. At the point $(x_c,1)$ we have 
more complicated behaviour, and this point is called a tricritical point.
\vspace{-0.9in}
\begin{figure}[ht!]
\begin{center} 
\includegraphics[width=8cm]{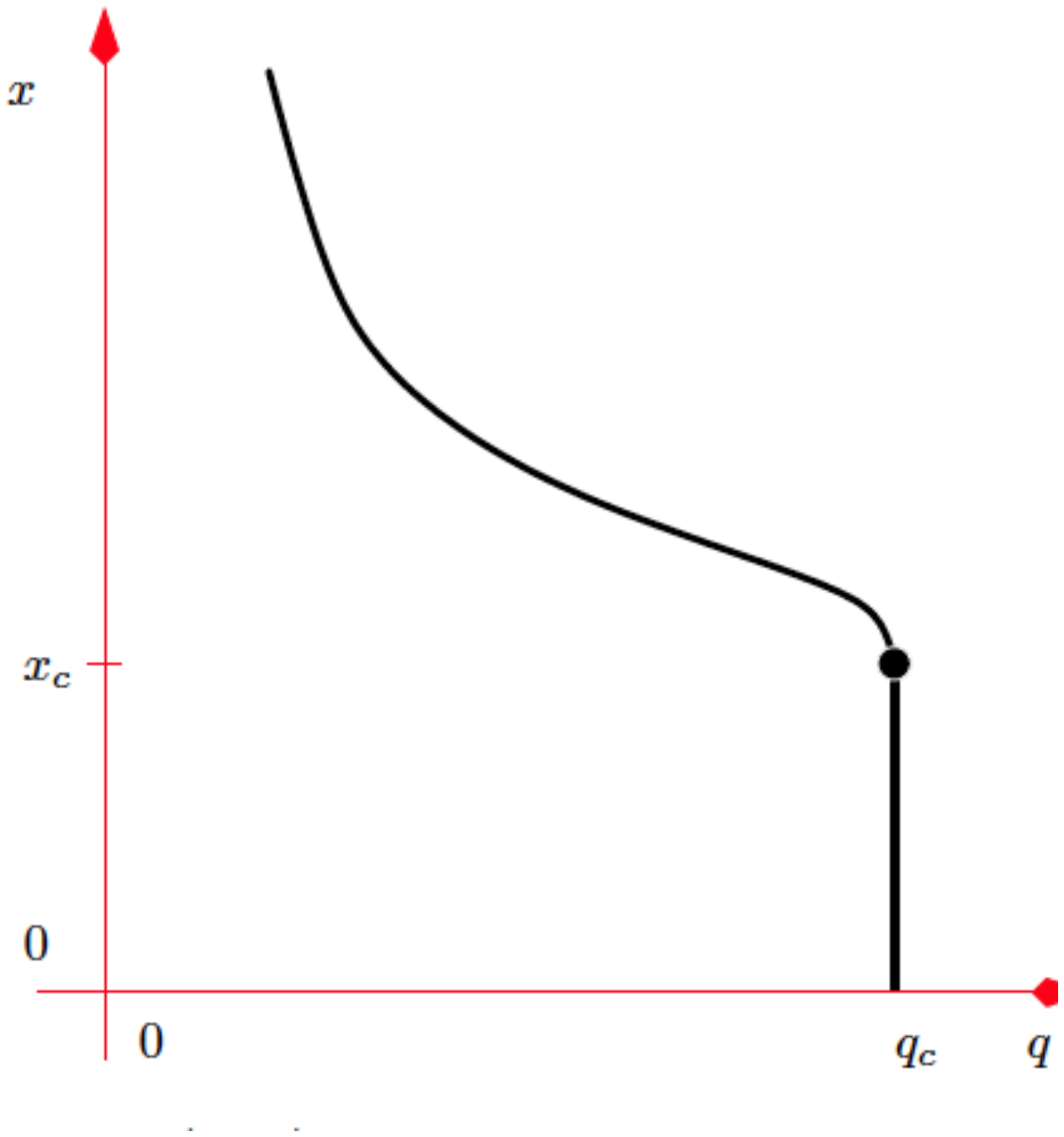}
\end{center}
\end{figure}

\vspace{-1.2in}

Around the point $(x_c,1)$ we expect tricritical behaviour, so that
\index{tricritical!scaling}
     \begin{equation}\label{nat1}
       \mathcal{P}^{(sing)}(x,q) \sim (1-q)^\theta F\left ((x_c-x)(1-q)^{-\phi}\right ) \,\, {\rm as} \,\,\, (x,q) \to (x_c,1^-).
     \end{equation}
     Here the superscript $(sing)$ means the singular part. There is an additional, additive part that is regular in the neighbourhood of $(x_c,1).$
 
    For self-avoiding polygons, in a series of papers, Richard and co-authors \cite{cr1, cr2, cr3} have provided abundant evidence (but no proof) for the surprisingly strong conjecture that
      \begin{equation}\label{nat2}
          F(s) = -\frac{1}{2\pi} \log\mbox{Ai} \left( \frac{\pi}{x_c} 
             \left( 4A_0\right)^\frac{2}{3} s \right)+C(q).
\end{equation}
Here $\mbox{Ai}(x)$ is the Airy function, and $C(q)$ is a function, independent of $x$, that arises as a constant of integration. 
Then equation (\ref{nat2}) implies that the exponents introduced in equation (\ref{nat1}) are $\phi = 2/3$ and $\theta=1.$ Here $A_0$ is a constant known only numerically.


In the next section we give the proof due to Duminil-Copin and Smirnov of the exact growth constant for the honeycomb lattice. In the following section we give three examples of applications of SAWs to other areas of science, and in the conclusion we give more detail of recent developments that we hope point the way to future breakthroughs.

\section{The honeycomb lattice}\label{sec:DC-S}
As mentioned above, a 
 breakthrough was achieved in 2010, when Duminil-Copin and
Smirnov \cite{DCS10}  proved that the growth constant on the honeycomb
lattice is $\mu=\sqrt{2+\sqrt 2}$, as predicted by Nienhuis \cite{N82}, using compelling physical arguments from conformal field theory, 30 years
previously.
The argument is, in hindsight, so simple, and the result so important, that we sketch it
here. 
\begin{figure}[htb]
\scalebox{0.5}{
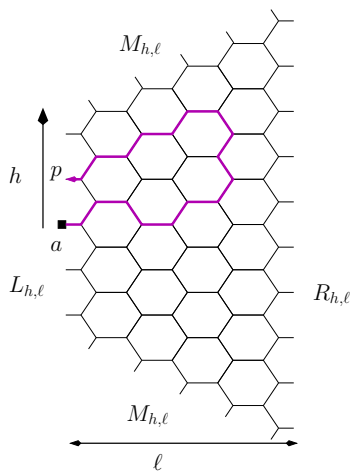}
\caption{A trapezoid $\cT$ on the honeycomb lattice.}
\label{fig:hexa}
\end{figure}

\begin{figure}[htb]
\scalebox{0.5}{
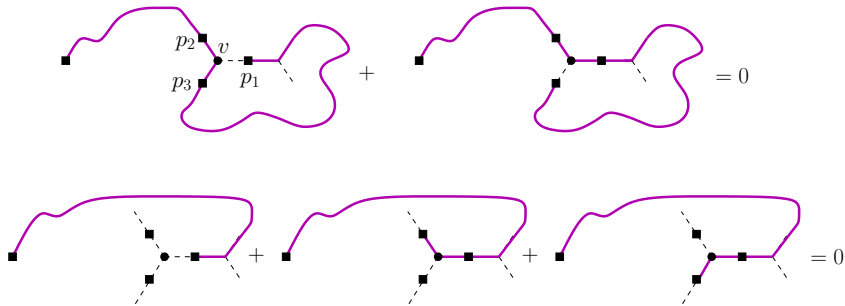}
\caption{A very local proof of the local identity~\eqref{id-DC-S}.}
\label{fig:local}
\end{figure}

We  consider SAWs that start from a point $a$
located on the left side of a trapezoid $\mathcal T$ of width $\ell$ and height
$h$, as shown in Figure~\ref{fig:hexa}. For $p$ a mid-edge of $\mathcal
T$, 
let $F(p)$ be the  generating function of
SAW $w$ that end at $p$, weighted by the number of vertices $\vv(w)$
and the  number of 
turns $T(w)$ (a left turn counts $+1,$ a right turn $-1$):
\begin{equation*}\label{DCS1}
F(p)\equiv F(p;x,\alpha):= \sum_{w: a \leadsto p
} x^{\vv(w)} e^{i \alpha T(w)}.
\end{equation*}
For instance, the walk of Figure~\ref{fig:hexa} visits 17 vertices, makes 10
left turns and 7 right turns, so that its contribution to $F(p)$
is $x^{17} e^{3i\alpha}$. Then, if $v$ is any vertex of $\mathcal T$
and $p_1, p_2, p_3$ are the three mid-edges adjacent to it, the following local
identity holds:
\beq\label{id-DC-S}
(p_1-v)F(p_1) + (p_2-v) F(p_2)+ (p_3-v) F(p_3)=0,
\eeq
provided  $x=x_c:=1/\sqrt{2+\sqrt 2},$ which is the reciprocal of the conjectured
growth constant, and 
$\alpha=-5\pi/24$.
(We consider  that the honeycomb lattice is embedded in the
complex plane $\cs$, so that $p_i-v$ is a complex number). This identity
is easily proved by grouping as pairs or triplets the SAWs that contribute
to its left-hand side, as depicted in Figure~\ref{fig:local}. One then checks
that the contribution of each group is zero.

If we now sum~\eqref{id-DC-S} over all vertices $v$ of $\cT,$ then  due to
the terms $(p_i-v)$,  all terms $F(p)$ such that $p$ is not a mid-edge
of the border disappear. After a few more reductions based on
symmetries, one is left with
$$
\left(\cos \frac {3\pi} 8\right) \, L_{h, \ell}(x_c) + \frac 1 {\sqrt{2}}\,  M_{h,\ell}(x_c)+
R_{h,\ell}(x_c)=1,
$$
where $L_{h, \ell}(x)$
(resp. $R_{h,\ell}(x)$, $M_{h,\ell}(x)$) is the generating function of
SAWs $w$ that end on the left side (resp. right side, top or bottom) of
$\mathcal T$, weighted by the number of vertices $\vv(w)$. 

%
By letting $h$ and then $\ell$ tend to infinity, Duminil-Copin and
Smirnov derived from this identity that the \gf\ of SAWs diverges
at $x_c$, but converges when $x<x_c$. This means that its radius of
convergence is $x_c$, so that the growth constant is $1/x_c=
\sqrt{2+\sqrt 2}$. 

Unfortunately these ideas do not generalise to SAW on the square or triangular lattices, for which we only have accurate numerical estimates for the growth constant $\mu.$
\section{Applications}\label{sec:apps}
One reason that SAWs and SAPs are so extensively studied, apart from their intrinsic mathematical interest, is that they model many problems that arise in other fields.
The first such example we will consider extends the proof given above to the situation where the SAW can interact with a surface. The second example considers SAWs crossing a square, with application to telecommunication networks, and the third example models some recent biological experiments where strands of DNA (a polymer) are pulled from a wall with optical tweezers.

\subsection{Walks attached to a surface}

The interaction of polymers with a surface is scientifically and industrially an important phenomenon. A common example is the adherence of paint to a surface, clearly an industrial process of considerable significance. To model such phenomena requires the inclusion of an interaction term between the polymer and the surface. To achieve this, we add a weight 
$y$ to vertices in the surface, as shown in Figure \ref{fig:weights} below. In physics terms, $y =\e^{-\epsilon/k_BT}$  where $\epsilon$ is
the energy associated with a surface vertex, $T$ is the absolute
temperature and $k_B$ is Boltzmann's constant. It is known that the growth constant $\mu = 1/x_c$ for such walks is the same as for the bulk case.

Let $c_n^+(i)$ be the number of half-plane walks of $n$-steps, with
$i$ monomers in the surface, and define the partition function (or \gf) as 
$$
C_n^+(y) = \sum_{i=0}^n c_n^+(i)y^i.
$$
 If $y$ is large, the polymer adsorbs onto the surface, while 
if $y$ is small, the walk is repelled by the surface. 

\begin{Proposition}\label{prop:half-space}
For $y>0$, 
$$ 
\mu(y):= \lim_{n  \to \infty} C^+_n(y)^{1/n} 
$$ 
exists and is finite. It is a log-convex, 
non-decreasing function of $\log y$, and therefore continuous
and almost everywhere differentiable.

For $0<y\le 1$, 
$$
\mu(y)=\mu(1)\equiv \mu.
$$
Moreover, for any $y>0$,
$$
\mu(y) \ge \max (\mu,\sqrt y ).
$$
This behaviour implies the existence of a critical value $\y$, with 
$1\le \y \le \mu^2$, which delineates the transition from the {\em
  desorbed} phase to the {\em adsorbed} phase:
$$
\mu(y) \left\{
  \begin{array}{ll}
    = \mu & \hbox{ if } y\le \y ,\\
>\mu & \hbox{ if } y >\y .
  \end{array}
\right.
$$
\end{Proposition}

In 1995 Batchelor and Yung~\cite{BY95} extended Nienhuis's \cite{N82} 
work to the adsorption problem just described, and making similar
assumptions to Nienhuis conjectured the value of the critical surface
fugacity for the  honeycomb lattice SAW model, to be $y_c =  1+\sqrt{2}.$
 In 2012 this was proved by Beaton, Bousquet-M\'elou, de Gier, Duminil-Copin and Guttmann \cite{BBdDCG12}, and here we will sketch their proof.

Take the same trapezoid as above, now called $S_{T,L}$, and add weights to the vertices on the $\beta$ boundary, as shown in bold in the figure below:
\begin{figure}[h]
\includegraphics[scale=0.8]{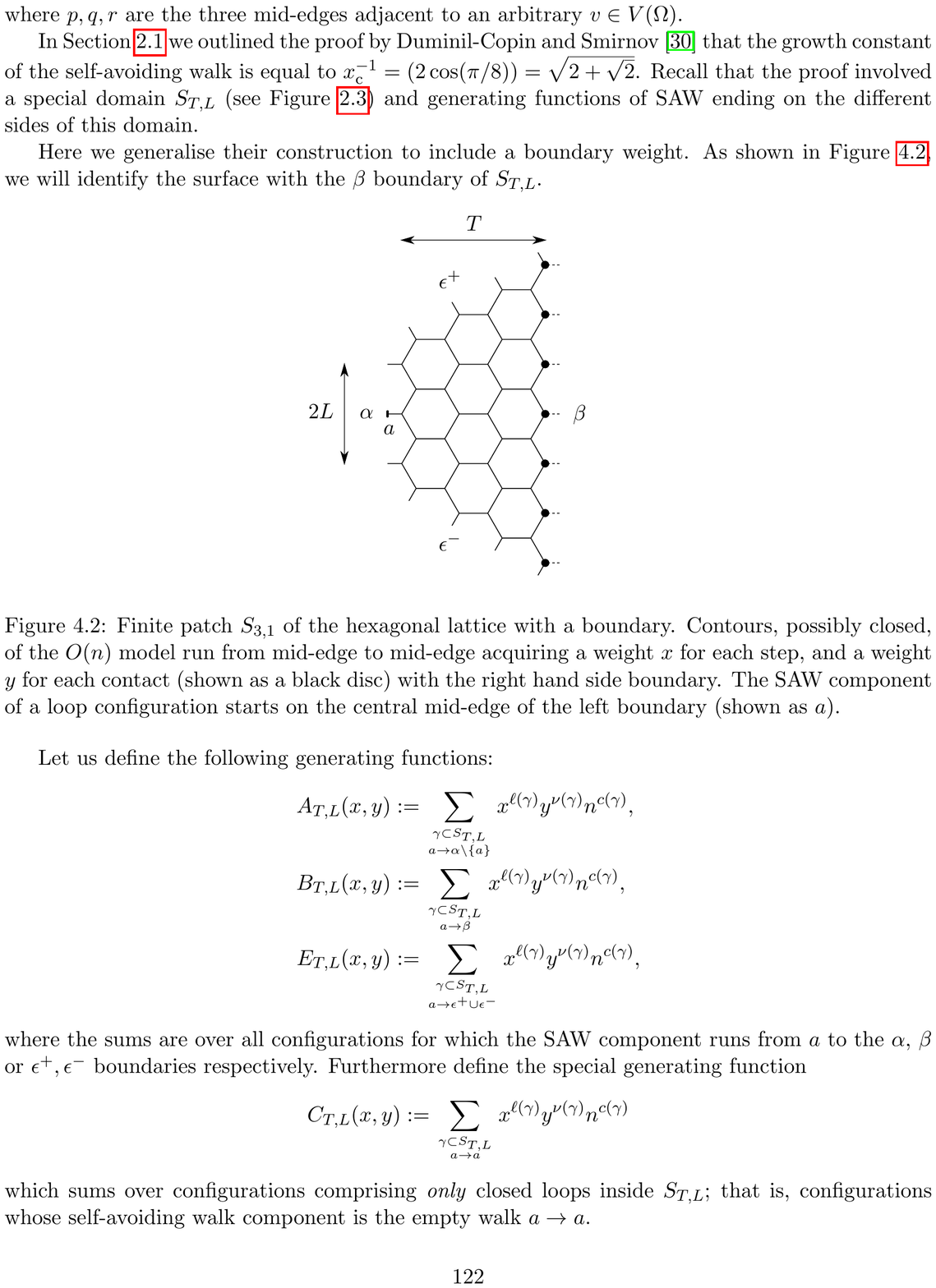}
\caption{Trapezoidal domain $S_{T,L}$ with vertices on the right-hand side wall, shown in bold, carrying a weight $y.$}
\label{fig:weights}
\end{figure}
Then we find the corresponding identity between generating functions, with $y^*=1+\sqrt{2},$ to be
{\small \[1=\cos\left(\frac{3\pi}{8}\right)A_{T,L}(\x,y) + \cos\left(\frac{\pi}{4}\right)E_{T,L}(\x,y) + \frac{y^*-y}{y(y^*-1)}B_{T,L}(\x,y)\]}
\noindent where $y$ is conjugate to the number of visits to the $\beta$ boundary. The generating functions $A_{T,L},$ $B_{T,L}$ and $E_{T,L}$ are two-variable generalisations of those defined in the previous section. To prove the conjecture we need to show that $y_c = y^*.$

It is safe to take $L\to\infty$ so that the trapezoid becomes a strip. The identity then becomes
\[1=\cos\left(\frac{3\pi}{8}\right)A_{T}(\x,y) + \cos\left(\frac{\pi}{4}\right)E_{T}(\x,y) + \frac{y^*-y}{y(y^*-1)}B_{T}(\x,y).\]

It is then straightforward to prove that\\
(i) $E_T(\x,y)=0$ for $0\leq y<y^*,$\\
(ii) $\y\geq y^*,$\\
(iii) $\lim_{T \to \infty}  A_T(\x,y) = A(\x,y) = A(\x)$ is {constant} for $0\leq y<y^*.$

If we now write
\[\cos(3\pi/8)A(\x,y) = 1-\delta\]
then the above identity reduces to
\[B(\x,y) = \lim_{T \to \infty} B_T(\x,y) = \frac{\delta y(y^*-1)}{y^*-y}\]
and in particular
\[B(\x,1) = \delta.\]

\begin{prop}
If $\delta=0$ then $\y=y^*.$
\end{prop}

The proof uses a decomposition of $A$ walks in a strip of width $T$ into $B$ walks in that same strip, and gives rise to an inequality.  In particular, for $y<\y = \lim_{T \to \infty} y_T$,
$$
0\le  \alpha \x  + \frac 1 {B_{T}(\x ,1)} \frac{y^*-y}{y(y^*-1)}.
$$
If $B_T(\x ,1)$ tends to $0,$ 
  this forces 
$y^*\ge \y ,$
otherwise the right-hand side would become arbitrarily large in
modulus and negative as $T\to \infty$ for $y^*<y<\y $.

Together with $y^* \le \y,$ this establishes $\y=\ys =1+\sqrt 2$ and
completes the proof of the proposition. 

The proof that $\delta=0$ is complicated, and unlike most other proofs we have given is almost totally probabilistic. It is unrealistic to give any details, but in essence one
first uses renewal theory to show that $\delta^{-1}$ is the expected height of an {\em irreducible bridge,} which is a SAW that crosses the strip from left to right, and cannot be expressed as the concatenation of two or more smaller such bridges. Next one
shows that, for irreducible bridges, $\mathbb{E}[width]<\infty$ implies that $  \mathbb{E}[height]<\infty.$
Finally one shows that the assumption that $\mathbb{E}[height]<\infty$  leads to a contradiction, from which the desired result that $\delta=0$ readily follows.

\subsection{Walks crossing a square}
Some years ago I was asked by a telecommunications engineer to  help him with the following problem: His company had a square grid of nodes, connected by wires, and phone-calls could  be routed from the bottom left-hand corner to the top right-hand corner of the grid. He wished to know how many such routes there were, as this determined the carrying capacity of the network.

After some discussion we agreed that this was simply the question {\em how many distinct SAWs  are there on a square grid of side-length $L$ originating at $(0,0)$ and ending at $(L,L)?$} The problem as stated was first considered by Knuth \cite{K76} in 1976, who gave a Monte Carlo estimate for the number of paths for $L = 10,$ a result we now know exactly. The problem was generalised by Whittington and the author \cite{WG90} to include a weight $x$ associated with each step of the walk. This gives rise
to a canonical model of a phase transition. For $x < 1/\mu$ the average length of
a SAW grows as $L,$ while for $x > 1/\mu$ it grows as $L^2.$ Here $\mu$ is the growth
constant of unconstrained SAWs on the square lattice, defined above. For $x = 1/\mu$  numerical
evidence, but no proof, was given that the average walk length grows as $L^{4/3}.$ 
Let $c_n(L)$ denote the number of walks of length $n.$ Clearly $c_n(L)=0$ for $n < 2L.$ We denote the generating function by $C_L(x) := \sum_n c_n(L) x^n.$ The answer to the original question is $\sum_n c_n(L).$

Subsequently,  Madras  \cite{M95} proved a number of relevant results. In fact, 
most of Madras's results were proved for the more general $d$-dimensional hypercubic lattice, 
but here we will quote them in the more restricted two-dimensional setting. 

\begin{Theorem}
The following limits,
$$
\mu_1(x) := \lim_{L \to \infty} C_L(x)^{1/L} \quad \hbox{ and }
\quad 
 \mu_2(x) := \lim_{L \to \infty} C_L(x)^{1/L^2},
$$
are well-defined in $\rs \cup \{+\infty\}$.

More precisely, 
\begin{itemize}
\item [$(i)$] $\mu_1(x)$ 
is finite for $0 < x \le 1/\mu$, and is infinite for
$x > 1/\mu.$ Moreover, $0 < \mu_1(x) < 1$ for $0 < x < 1/\mu$ and
$\mu_1(1/\mu) = 1.$
\item [$(ii)$] 
 $\mu_2(x)$ is finite for all $x > 0.$
Moreover, $\mu_2(x) = 1$ for $0 < x \le 1/\mu$ and $\mu_2(x) >
1$ for $x > 1/\mu.$ 
\end{itemize}
\end{Theorem}

In \cite{WG90} the existence of the limit $\mu_2(x)$ was proved, and in addition
upper and lower bounds on $\mu_2(x)$ were established.

The average length of a (weighted) walk is defined to be 
\begin{equation}\label{eq3}
\<n(x,L)\> := \sum_n n c_n(L)x^n/\sum_n  c_n(L)x^n. 
	\end{equation}

We define $\Theta(x)$ as follows:
Let  $a(x)$ and $b(x)$ be  two functions of some variable $x$. We
write  that $a(x) = \Theta( b(x))$ as $x\rightarrow x_0$ if there
exist two positive constants $\kappa_1$ and $\kappa_2$ such that,
for $x$ sufficiently close to $x_0$,
$$
\kappa_1\; b(x) \le a(x) \le \kappa_2\; b(x).
$$

\begin{Theorem}
 For $0 < x < 1/\mu$, we have that $\<n(x,L)\>= \Theta(L)$ as
 $L\rightarrow  \infty$, while  for
 $x > 1/\mu$, we have 
$\<n(x,L)\>=\Theta(L^2)$.
\end{Theorem}

In \cite{WG90} it was proved that $\<n(1,L)\>=\Theta(L^2)$.
The situation at $x = 1/\mu$ is unknown. In \cite{BGJ05} we gave 
compelling numerical evidence that in fact $\<n(1/\mu,L)\>
=\Theta(L^{1/\nu})$ , where $\nu = 3/4$, in 
accordance with an intuitive suggestion of Madras in \cite{M95}.

\begin{Theorem}
For $x > 0$, define $f_1(x) = \log \mu_1(x)$ and
 $f_2(x) = \log \mu_2(x).$
\begin{itemize}
\item [$(i)$]
 The function $f_1$ is a strictly increasing, negative-valued convex
 function of 
$\log x$ for $0 < x < 1/\mu.$ 
\item [$(ii)$]
The function $f_2$ is a strictly increasing, convex function of
$\log x$ for $x > 1/\mu,$ and satisfies $0 < f_2(x) \le \log \mu +
\log x.$
\end{itemize}
\end{Theorem}

Some, but not all of the above results were previously proved in \cite{WG90}, but these three theorems
elegantly capture all that is rigorously known. In \cite{BGJ05} an extensive numerical study was described, including exact enumerations up to squares of side 19. For the largest square there are exactly 1 523 344 971 704 879 993 080 742 810 319 229 690 899 454 255 323 294 555 776 029 866 737 355 060 592 877 569 255 844 distinct paths! The number of such paths, as we have seen, grows as $\lambda^{L^2}.$ In \cite{BGJ05} it was also proved that $1.628 < \lambda < 1.782$ and estimated that $\lambda = 1.744550 \pm 0.000005.$ 

\subsection{Walks in a rectangle}
In 2002 L N Trefethen \cite{T02} presented ten problems used in teaching numerical analysis at Oxford University. Problem number 10 asks: {\em A particle at the centre of a $10 \times 1$ rectangle undergoes Brownian motion (i.e., 2-D random walk with infinitesimal step lengths) till it hits the boundary. What is the probability that it hits at one of the ends rather than at one of the sides?}

Remarkably, it turns out \cite{BLWW04} that a closed form solution to this question can be found, {\em via} the theory of elliptic functions and invoking some results of Ramanujan on singular moduli. The result is $$p_e=\frac{2}{\pi}\arcsin\left ( (3 - 2\sqrt{2})^2(2+\sqrt{5})^2(\sqrt{10}-3)^2(5^{1/4}-\sqrt{2})^4 \right ) = 0.000000383758797925 \ldots .$$ In \cite{GK12}, the question as to the corresponding problem where Brownian motion is replaced by the scaling limit of a self-avoiding walk (SAW) is treated.  In fact, they addressed the equivalent question, which is the ratio of the probability that the walk hits the end ($p_e$) to the probability that it hits a side. This is just $$\frac{p_e}{1-p_e}$$ which has the numerical value $0.00000038375894519599411176841999126970034234598936\ldots$ for the random walk case. 

By using conformal mappings, it was shown in \cite{GK12} how to solve this problem for both Brownian motion and for the scaling limit of SAW, assuming that this is describable by $SLE_{8/3},$ as discussed in Sections 5.1 and 5.2 below.

Let $D$ be a bounded, simply connected domain in the complex plane containing $0.$ We are interested in paths in $D$ starting at $0$ and ending on the boundary of the domain. For Brownian motion,  the distribution of the end-point is harmonic measure. Let $h_D(z)$ denote the density with respect to arc length (often called the Poisson kernel). If $f$ is a conformal map on $D$ that fixes the origin, and such that the boundary of $f(D)$ is also piecewise smooth, then the conformal invariance of Brownian motion implies that the density for harmonic measure on the boundary of $f(D)$ is related to the boundary of $D$ by 
\begin{equation}\label{cirw}
h_D(z)=|f'(z)|h_{f(D)}(f(z)).
\end{equation}
For SAW, Lawler, Schramm and Werner \cite{LSW04} predicted that the corresponding density of the probability measure $\rho(z)$  transforms under conformal maps as
\begin{equation}\label{cisaw}
\rho_D(z)=c|f'(z)|^{\bexp}\rho_{f(D)}(f(z)),
\end{equation}
where $\bexp=5/8$ and the constant $c$ is required to ensure that $\rho_D(z)$ is a probability density. Note that $\bexp$ is related to $\kappa$ by $\bexp=3/\kappa - 1/2,$ so $\bexp=5/8$ corresponds to $\kappa=8/3.$ If one starts the random walk or the SAW at the center of a disc, then the hitting density on the circle will be uniform.  So the above equations determine the hitting density for any simply connected domain. 

The map $f$ between the upper half-plane and a rectangle (see figure below), is given by a Schwarz-Christoffel transformation. For $\alpha > 1,$ let
$$f(z) = \int_0^z \frac{d\xi}{\sqrt{1-\xi^2}\sqrt{\alpha^2 - \xi^2}}.$$
The rectangle has one edge along the real axis and $0$ is a midpoint of this side. So the corners
can be written as $\pm a/2$ and $ic \pm a/2$ where $a, c > 0$ are the length of the horizontal and vertical
edges, respectively. We have
$$f(1) = a/2, \,\, f(-1)=-a/2, \,\, f(\alpha) = a/2 + ic, \,\, f(-\alpha) = -a/2+ic, \,\, f(0) =0.$$

\begin{center}
\begin{tikzpicture}[xscale=0.8]
  \begin{scope}
    \node at (0,5) {$w$ plane};
    \filldraw[line width=2pt,fill=black!5]
    (-2,0)
    node[below left] {$C$}
    node[below left=10pt] {$-\frac{a}{2}$}
    -- (0,0)
    node[below left] {$D$}
    node[below right] {$0$}
    -- (2,0)
    node[below right] {$E$}
    node[below right=10pt] {$\frac{a}{2}$}
    -- (2,2)
    node[above right] {$F$}
    node[above right=10pt] {$\frac{a}{2} + i c$}
    -- (0,2)
    node[above right] {$G$}
    node[above left] {$A$}
    -- (-2,2)
    node[above left] {$B$}
    node[above left=10pt] {$\frac{a}{2} - i c$}
    -- cycle;
    \draw (0, -1) -- (0, 3.8) node[right] {$v$};
    \draw (-4,0) -- (4,0) node[above] {$u$};
  \end{scope}
  \begin{scope}[xshift=11cm]
    \node at (0,5) {$z$ plane};
    \fill[black!5] (-4,0) -- (4, 0) -- (4, 4) -- (-4,4);
    \draw (0, -1) -- (0, 3.8) node[right] {$y$};

    \filldraw[line width=2pt,-stealth',shorten >=-5pt] (-3.5, 0)
    -- (-3,0) node[above] {$A'$};
    \filldraw[line width=2pt] (-3,0)
    -- (-2,0) circle (2pt)
    node[above] {$B'$}
    node[below] {$-\alpha$};
    \filldraw[line width=2pt] (-2,0)
    -- (-1,0) circle (2pt)
    node[above] {$C'$}
    node[below] {$-1$}
    -- (-0,0) circle (2pt)
    node[above right] {$D'$}
    node[below right] {$0$}
    -- (1,0) circle (2pt)
    node[above] {$E'$}
    node[below] {$1$}
    -- (2,0) circle (2pt)
    node[above] {$F'$}
    node[below] {$\phantom{-}\alpha$};
    \filldraw[line width=2pt,-stealth',shorten >=-5pt] (2,0)
    -- (3,0) node[above] {$G'$};
    \draw[line width=2pt] (3,0) -- (3.5, 0) node[above] {$x$};
  \end{scope}
  \draw[->] (6.8,3) .. controls (5.8,3.3) and (5.2,3.3) .. (4.2,3)
  node[midway,above] {$w = f(z)$};
\end{tikzpicture}
\end{center}
So
$$ a=\int_{-1}^1 \frac{dx}{\sqrt{1-x^2}\sqrt{\alpha^2 - x^2}}, \,\,\, c =\int_{1}^\alpha \frac{dx}{\sqrt{x^2-1}\sqrt{\alpha^2 - x^2}}.$$ 
We note that $$ a =\frac{2}{\alpha}{\bf K} \left ( \frac{1}{\alpha} \right ), \,\, c=\frac{1}{\alpha}{\bf K} \left ( \frac{\sqrt{\alpha^2-1}}{\alpha} \right ), \,\, \alpha > 1.$$
Here ${\bf K}(x)$ is the complete elliptic integral of the first kind. By dilation invariance we only need concern ourselves with the aspect ratio $a/c.$ So given an aspect ratio
$r$, we have to find $\alpha$ such that

$$ r = \frac{\int_{-1}^1 (1-x^2)^{-1/2}(\alpha^2 - x^2)^{-1/2} dx}{\int_{1}^{\alpha} (x^2-1)^{-1/2}(\alpha^2 - x^2)^{-1/2} dx} = \frac{2 {\bf K} \left ( \frac{1}{\alpha} \right )}{{\bf K} \left ( \frac{\sqrt{\alpha^2-1}}{\alpha} \right )}.$$

Setting $k=1/\alpha,$ then $c$ can be written ${\bf K}(k')={\bf K}'(k).$ Therefore $$\frac{c}{a}=\frac{1}{r}=\frac{{\bf K}(1/\alpha')}{2{\bf K}(1/\alpha)}=\frac{{\bf K}'(1/\alpha)}{2{\bf K}(1/\alpha)}.$$ Now the nome $q$  is defined\footnote{Amramowitz and Stegun, p591 formula 17.3.17} by $q=\exp(-\pi{\bf K}'/{\bf K}),$ so $$r=\frac{{2\bf K}(1/\alpha)}{{\bf K'}(1/\alpha)},$$ and $$e^{-2\pi/ r} = q.$$ Then we have
$$\frac{1}{\alpha}= \left ( \frac{\theta_2(q)}{\theta_3(q)} \right )^2,$$ so 
$$\sqrt{\alpha} = \frac{\theta_3(e^{-2\pi/r})}{\theta_2(e^{-2\pi/r})},$$
where $\theta_j(q)=\theta_j(0,q)$ is the Jacobi theta function.
Evaluating this with any computer algebraic package gives the required value of $\alpha$ for any $r \ge 1$ instantly.

Alternatively, we can achieve good  accuracy by expanding the
ratio of the above integrals around $\alpha=1.$ This gives
$$r=\frac{1}{\pi}\left ( 4\log(2\sqrt{2})-2\log(\alpha-1)+(\alpha-1)-\frac{3}{8}(\alpha-1)^2+\frac{5}{24}(\alpha-1)^3+O(\alpha-1)^4 \right ).$$
Solving this numerically for $r=10$ gives $\alpha=1.00000120561454706472212\ldots.$
Reverting this equation, retaining the first two terms gives
\begin{equation} \label{alfa}
\alpha = 1+8 e^{-\pi r/2}+32 e^{-\pi r}+O( e^{-3\pi r/2}),
\end{equation}
 which, for $r=10$ gives 19 significant digits. 

To proceed further, we need the preimage of the center of the rectangle, which by symmetry is on the imaginary axis, so write it as $id.$
Either by symmetry arguments, or from the solution of the mapping equation, we readily find \cite{GK12} that $d^2 = \alpha.$

For the random walk and the SAW in the half plane starting at $id,$ it follows from (\ref{cisaw}) and (\ref{cirw}) that the (unnormalized)
hitting density along the real axis is $(x^2 + d^2)^{-\bexp} = (x^2 + \alpha)^{-\bexp}$. So letting $\rho_R$ denote the hitting density for a rectangle with the walk starting at the centre, we have by (\ref{cisaw}) and (\ref{cirw}),
\begin{equation}\label{rh}
(x^2 + \alpha)^{-\bexp} \propto |f'(x)|^\bexp \rho_R(f(z)).
\end{equation}

We require the ratio of the integral of $\rho(z)$ along a vertical edge to the integral along a horizontal edge,
$$\frac{\int_{0}^{c} \rho_R(a/2+iy) dy}{\int_{-a/2}^{a/2} \rho_R(x) dx}.$$
By a change of variable, setting $u = f^{-1}(x)$ in the denominator and $u=f^{-1}(a/2+iy)$ in the numerator, this becomes
 $$ \frac{\int_{1}^{\alpha} \rho_R(f(u))|f'(u)| du}{\int_{-1}^{1} \rho_R(f(u))|f'(u)| du}.$$
Since $f'(u) = (1-u^2)^{-1/2}(\alpha^2 - u^2)^{-1/2},$ the ratio of probabilities of a first hit on the vertical side to a first hit on the horizontal side, $R(\alpha,\bexp)$ is
\begin{equation}\label{rat}
R(\alpha,\bexp) = \frac {\int_{1}^{\alpha} (u^2+\alpha)^{-\bexp}(u^2-1)^{(\bexp-1)/2}(\alpha^2 - u^2)^{(\bexp-1)/2} du}{\int_{-1}^{1} (u^2+\alpha)^{-\bexp}(1-u^2)^{(\bexp-1)/2}(\alpha^2 - u^2)^{(\bexp-1)/2} du}.
\end{equation}
Note that this result should hold for $\bexp \in (1/4,1],$ so applies more broadly than just to SAWs.

The Brownian motion case corresponds to $\bexp=1.$ The integrals in (\ref{rat}) greatly simplify, giving
$$R(\alpha,1) = \frac{\arctan(\sqrt{\alpha})-\arctan(1/\sqrt{\alpha})}{2\arctan(1/\sqrt{\alpha})}.$$

It is straightforward to calculate the asymptotic expansion of both the numerator and denominator, and hence their ratio. In this way we find
$$R(\alpha,1) = \frac{8}{\pi}e^{-\pi r/2}  + \frac{64}{\pi^2}e^{-\pi r}+ O\left (e^{-3\pi r/2}\right ).$$
For $r=10$ this evaluates to $3.8375894519594\ldots \times 10^{-7},$ which is correct to 13 significant digits. 

Unfortunately for $\bexp \ne 1$ one cannot evaluate the integrals exactly, but one can evaluate the relevant integrals numerically. For SAWs $\bexp = 5/8$ and one finds ${\tilde R}(10,5/8) \approx 6.682989935 \times 10^{-5},$ some 200 times larger than the corresponding result for SAWs.

As for the random walk case, asymptotics also provides us with reasonable accuracy. Referring to equation (\ref{rat}), and noting that $\alpha$ is very close to 1 when the aspect ratio is 10, the denominator integral can be accurately approximated by $$
\int_{-1}^{1} (u^2+1)^{-\bexp}(1-u^2)^{\bexp-1}du = \frac{\sqrt{\pi} \Gamma \left (\frac{\bexp}{2}\right )}{2\Gamma\left (\frac{\bexp}{2}+\frac{1}{2}\right )}.$$ 

The numerator integral has a very narrow range of integration, and so can be approximated by $$\frac{1}{2}
\int_{1}^{\alpha} (u-1)^{(\bexp-1)/2}(\alpha-u)^{(\bexp-1)/2}du.$$ If we set $u=1+t(\alpha-1)$, this becomes
$$\frac{1}{2}(\alpha-1)^\bexp\int_0^1[t(1-t)]^{(\bexp-1)/2}dt= \frac{2^{-\bexp-1}\sqrt{\pi}\Gamma\left (\frac{1+\bexp}{2}\right )}{\Gamma \left (1+\frac{\bexp}{2}\right )}(\alpha-1)^\bexp.$$

Combining the above results, and using the result derived above, $\alpha-1 \approx 8 e^{-\pi r/2},$ one finds, asymptotically
\begin{equation}
\label{finas}
{\tilde R}(r,\bexp) \approx \frac{2^{2\bexp} \Gamma \left ( \frac{1}{2}+\frac{\bexp}{2} \right )^2}{\Gamma\left ( 1+ \frac{\bexp}{2} \right ) \Gamma \left ( \frac{\bexp}{2} \right )} e^{-\pi \bexp r/2}.
\end{equation}
For SAW equation (\ref{finas}) reduces to $${\tilde R}(r,5/8) \approx 1.2263431442 e^{-5\pi r/16}.$$
For $r=10$ this gives ${\tilde R}(10,5/8) \approx 6.6824528 \times 10^{-5},$ which is accurate to 4 significant digits.

A rather elaborate calculation using Mellin transforms gives the next term in the asymptotic expansion of the aspect ratio as

$${\tilde R}(r,b)=\frac{2^{2b+1}\Lambda}{b} e^{-b\pi r/2}\left [ 1 + \frac{\Lambda 2^{b+1}} {b \sin\left ( \frac{\pi b}{2} \right )}e^{-b\pi r/2} + 4(b-1+2\Lambda)e^{-\pi r/2}+ O(e^{-b\pi r})\right ],$$ for $0 < b < 1,$ where $\Lambda = \left (\frac{\Gamma \left ( \frac{1+b}{2} \right ) }{\Gamma \left ( \frac{b}{2} \right ) }\right )^2.$ Evaluating this for $r=10, \,\, b= \frac{5}{8}$, we find $${\tilde R}(10,5/8)= 0.00006682989679\cdots$$ which is in error by 2 in the $8^{th}$ significant digit.

Note that the ratio for the scaling limit of SAWs in a $10 \times 1$ rectangle is  some two orders of magnitude larger than the corresponding result for Brownian motion. This can be qualitatively understood as SAWs having a greater tendency to go in straight lines than do random walks.

\subsection{Pulling a polymer from a wall}
During the past decade, force has been used as a thermodynamic variable 
to understand molecular interactions and their role in the structure of 
bio-molecules \cite{rief,busta,tskhovrebova}. By exerting a force in the picoNewton range, one aims to experimentally study and characterize the elastic, mechanical, 
structural and functional properties of bio-molecules \cite{busta1}. 

\begin{figure}[h]
\includegraphics[width=2.6in]{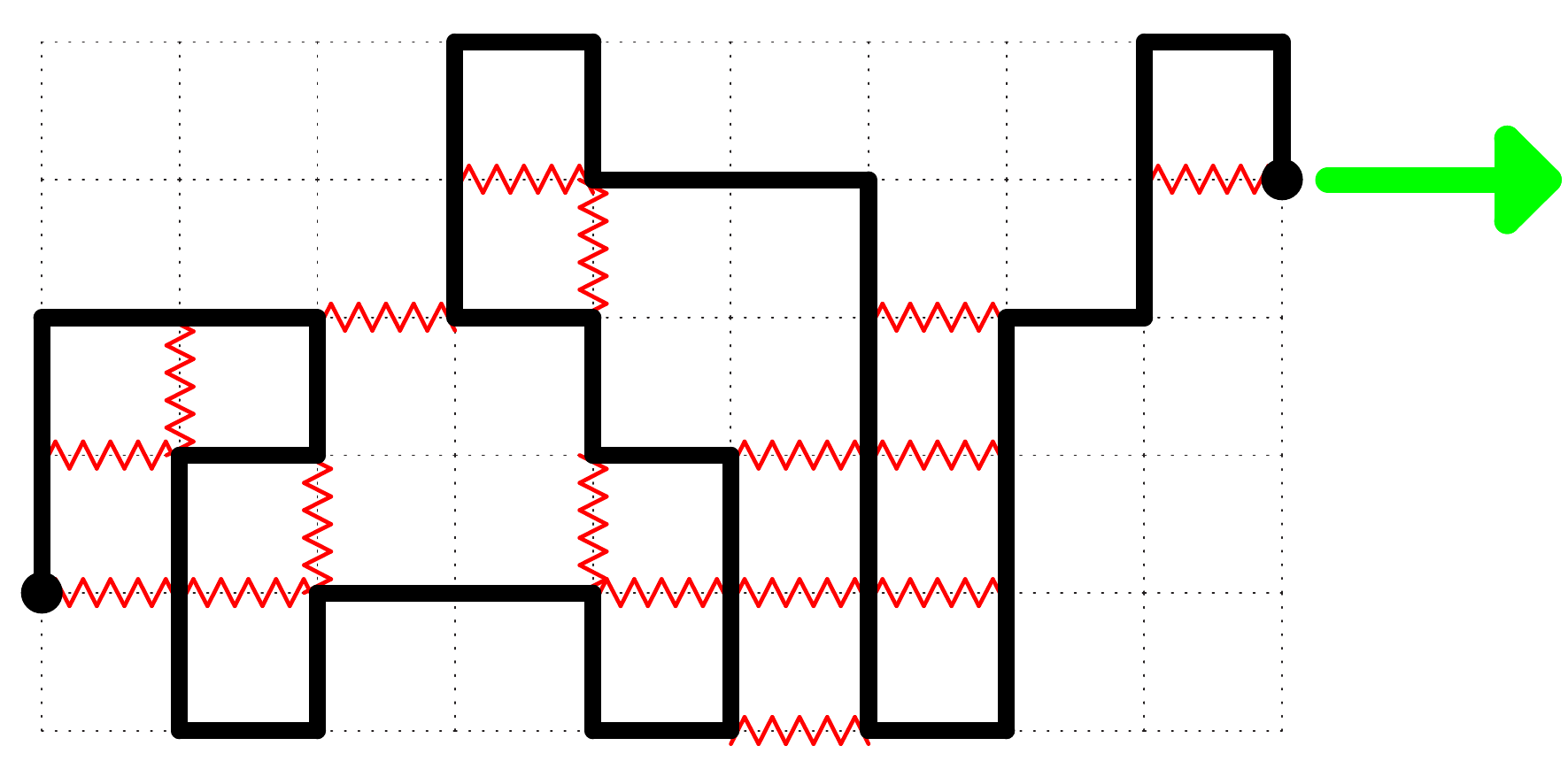}
\caption{\label{fig:model} An interacting self-avoiding walk on the square 
lattice with one end attached to a surface and subject to a pulling force 
at the other end. Each step of the walk connecting a pair of monomers 
is indicated by a thick solid line while interactions between non-bonded 
nearest neighbour monomers are indicated by jagged lines.}
\end{figure}

In \cite{KJJG07} SAWs were used to model the situation in which a polymer is attached to a surface and pulled from that surface by an applied force. The situation is shown in Figure \ref{fig:model}.
Interactions are introduced between neighbouring monomers on 
the lattice that are not adjacent along the chain. The pulling force is 
modeled by introducing an energy proportional to the $x$-component of 
the end-to-end distance.
One end of the polymer is attached to an impenetrable 
surface while the polymer 
is being pulled from the other end with a force acting along the 
$x$-axis.  

Boltzmann weights
$\omega=\exp(-\epsilon/\BC T$) and $u=\exp(-F/\BC T)$ conjugate
to the nearest neighbor interactions and force, respectively, were introduced, where
$\epsilon$ is the interaction energy,
$\BC$ is Boltzmann's constant, $T$ the temperature and $F$ the
applied force. For simplicity, we set $\epsilon=-1$ and $\BC=1$.
The relevant finite-length partition functions are
\begin{equation}
Z_N(F,T)  = \!\!\!\!\! \sum_{all \; walks} \!\!\!\!\!\!\! \omega^m u^x 
     \! =  \sum_{m,x} \! C(N,m,x)  \omega^m u^x,
\end{equation}
where $C(N,m,x)$ is the number of interacting SAWs of length $N$ 
having $m$ nearest neighbor contacts and whose end-points 
are a distance $x=x_N-x_0$ apart.
 The partition functions of the constant force ensemble, $Z_N(F,T)$, 
and constant distance ensemble, $Z_N(x,T)= \sum_{m} C(N,m,x) \omega^m$, are related by
$Z_N(F,T)  =  \sum_{x} Z_N(x,T) u^x$.
The free energies are evaluated from the partition functions 
\begin{eqnarray}
G(x)= -T \log Z_N(x,T) \; \; {\rm and } \; \;
G(F)= -T \log Z_N(F,T).
\end{eqnarray}
Here $\langle x \rangle = \frac{\partial G(F)}{\partial F}$ and 
$\langle F \rangle = \frac{\partial G(x)} {\partial x}$ are the 
control parameters of the constant force
and constant distance ensembles, respectively.

All possible conformations of the SAW were enumerated. 
The challenge facing
exact enumerations is to increase the chain length. Using direct
counting algorithms the time required to enumerate all the configurations
increases as $\mu^N$, where $\mu$ is the connective constant of  
the lattice  ($\mu \approx 2.638$ on the square lattice). 
So even with a rapid increase in computing power only a few
more terms can be obtained each decade.
In \cite{KJJG07} the number of  interacting SAWs was calculated using
transfer matrix techniques \cite{jensen}. Combined with parallel processing these
algorithms allowed  the enumerations to be extended to chain lengths up to 55 steps, roughly doubling the previously available enumerations.

\begin{figure}[t]
\includegraphics[width=2.6in]{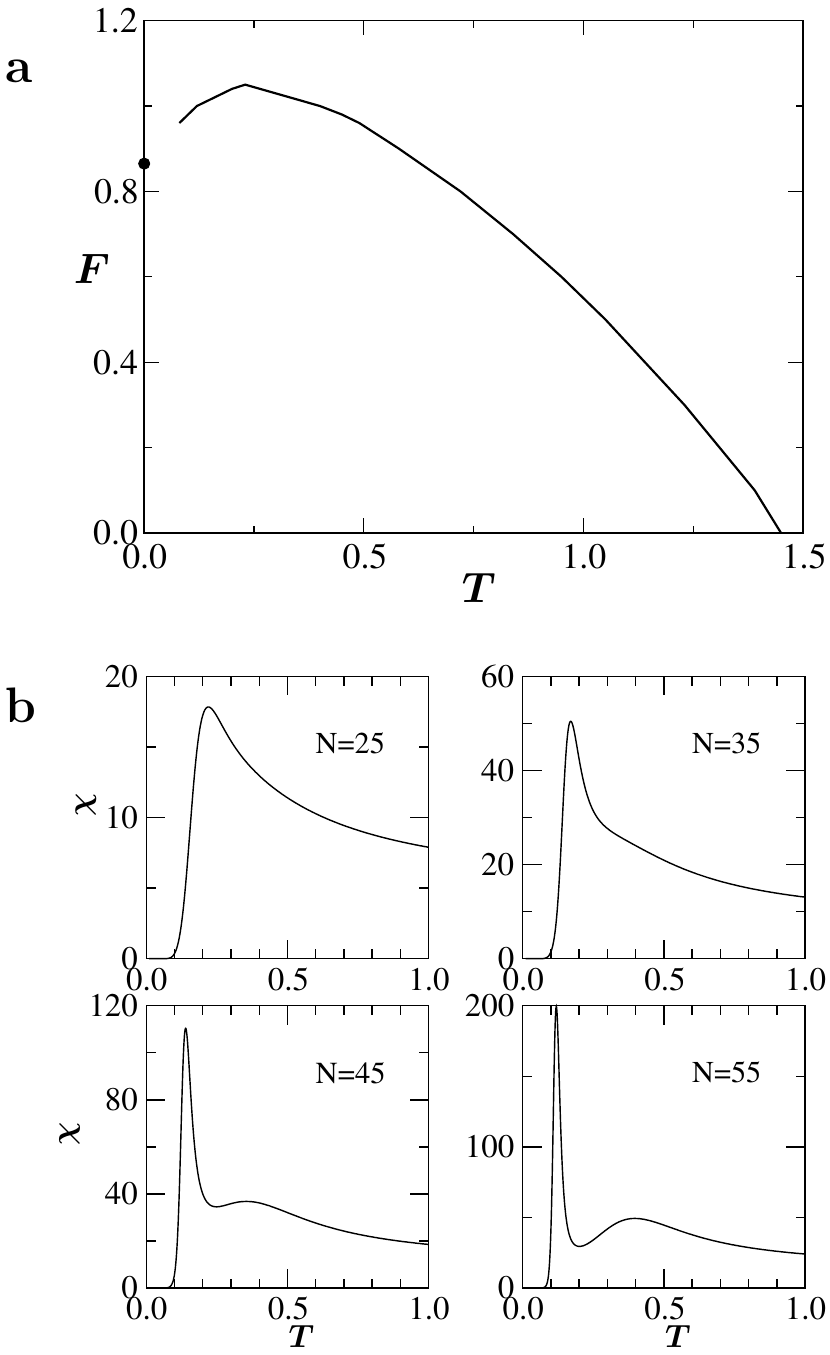}
\caption{\label{fig:PDflex} 
{\bf a} The force-temperature phase diagram for flexible chains.  
{\bf b} The fluctuations $\chi$ in the number of contacts 
vs. temperature at fixed force $F=1.0$ for various values of the chain length $N$.}
\end{figure}

In Fig.~\ref{fig:PDflex}a we show the force-temperature 
phase diagram for flexible chains. At low temperature and force the polymer
chain is in the collapsed state and as the temperature is increased (at fixed
force) the polymer chain undergoes a phase transition to an extended state.
The transition temperature as plotted in  Fig.~\ref{fig:PDflex}a was found 
(in the thermodynamic limit $N\to \infty$) by studying the reduced free energy per 
monomer. The most notable feature of  the phase-diagram is the
{\em re-entrant} behaviour--that is to say, the initial increase of force with temperature, prior to a decrease. 

The positive slope $dF_c/dT$ at $T=0$ confirms the existence of re-entrance
in the $F-T$ phase-diagram. The authors pointed out that the value of the transition temperature
obtained in the thermodynamic limit and the one  obtained from the fluctuations
in non-bonded nearest neighbors (which can be calculated exactly for
finite $N=55$) gives almost the same value (within error bars of $\pm0.01$).
The fluctuations are defined as $\chi=\langle m^2 \rangle -\langle m \rangle^2$,
with the $k$'th moment given by 
$$\langle m^k \rangle = 
\sum_{m,x} \! m^k C(N,m,x)  \omega^m u^x/\sum_{m,x} \! C(N,m,x)  \omega^m u^x.$$
In the panels of Fig.~\ref{fig:PDflex}b  the emergence of two peaks in 
the fluctuation curves with increasing $N$ at fixed force $F=1.0$ are shown. The twin-peaks
reflect the fact that in the re-entrant region as one increases $T$ (with $F$ fixed)
the polymer chain undergoes two phase transitions. The importance of powerful enumeration data is highlighted by the observation that the twin-peaks are
not apparent for small values of $N$. Many more details and comparison with experiments are given in \cite{KJJG07} -- our purpose here is just to show the applicability of SAWs to this problem.

\section{The scaling limit and SLE}
\subsection{The scaling limit}
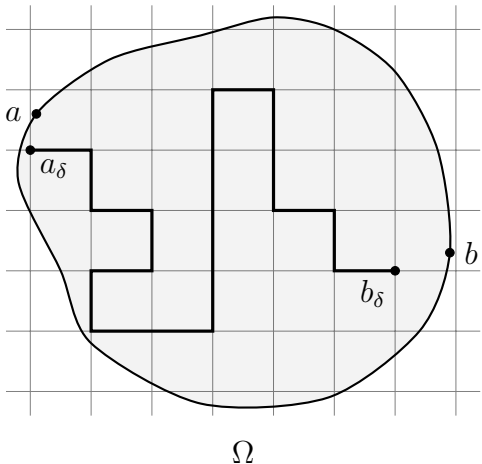
\begin{figure}\label{fig:scal-lim}
\centering
\begin{tikzpicture}[every node/.style={font=\Large},scale=0.8]
\draw[help lines,step=1cm] (0.6,0.6) grid (8.4,7.4);
\coordinate (A) at (1.1,5.6);
\coordinate (Adelta) at (1,5);
\coordinate (B) at (7.9,3.3);
\coordinate (Bdelta) at (7,3);
\draw (A) node [left=2pt] {$a$} [fill] circle (2pt);
\draw (Adelta) node [below right] {$a_{\delta}$} [fill] circle (2pt);
\draw (B) node [right=2pt] {$b$} [fill] circle (2pt);
\draw (Bdelta) node [below left] {$b_{\delta}$} [fill] circle (2pt);
\filldraw[thick,smooth cycle,fill opacity=0.05] plot coordinates {(0.8, 4.5) (A) (2.3,6.5) (3.8,6.9) (5,7.2) (6,7) (7,6.3) (7.7,5) (B) (7.4,2) (5.9,0.9) (3.8,0.8) (2,1.8) (1.5,3) };
\draw[very thick] (Adelta) -- ++(1,0) -- ++(0,-1) -- ++(1,0) -- ++(0,-1) -- ++(-1,0) -- ++(0,-1) -- ++(2,0) -- ++(0,4) -- ++(1,0) -- ++(0,-2) -- ++(1,0) -- ++(0,-1) -- (Bdelta);
\draw (4.5,0) node {$\Omega$};
\end{tikzpicture}
\caption{Discretisation of domain $\Omega.$}
\end{figure}

One topic we have failed to adequately address is the question of the {\em scaling limit} of SAW. An intuitive grasp of this concept can perhaps be gained by looking at the first two figures in this article. In the first figure, the effect of the lattice is clear. In the second figure, there is no obvious lattice, and indeed no way to tell that this is not a continuous curve. We formalise this notion as follows: Consider a smooth (enough) closed domain $\Omega,$ with an underlying square grid, with grid spacing $\delta$ as shown in Figure \ref{fig:scal-lim}. Denote by $\Omega_\delta$ that portion of the grid contained in $\Omega.$ Take two distinct points on the boundary of $\Omega$ labelled $a$ and $b.$
Now take the nearest lattice vertex to $a,$ and label it $a_\delta,$ and similarly $b_\delta$ is the label of the nearest lattice vertex to point $b.$ Consider the set of SAWs  $\omega(\Omega_\delta)$ on the finite domain $\Omega_\delta$ from $a_\delta$ to $b_\delta.$ Recall that $\delta > 0$ sets the scale of the grid. Now let $|\omega|$ be the length of a walk  $\omega_\delta \in \omega(\Omega_\delta),$ and weight the walk by $x^{|\omega|}.$ The reason for this is that the walks are of different lengths, making the uniform measure not particularly natural. (There is also a normalising factor, which for simplicity we ignore).

As we let $\delta \to 0$ we expect the behaviour of the walk to depend on the value of $x.$ For $x < x_c$ it is possible to prove that $\omega_{\delta}$ goes to a straight line as $\delta \to 0.$ (Strictly speaking it converges in distribution to a straight line, with fluctuations O$(\sqrt{\delta}).$)
For $x > x_c$ it is expected that $\omega_{\delta}$ becomes (again, in distribution) space-filling as $\delta \to 0.$ But at $x = x_c$ it is {\em conjectured} that $\omega_{\delta}$ becomes (in distribution) a random continuous curve, and is conformally invariant. This describes the scaling limit. If this conjecture is correct, a second, pivotal, conjecture by Lawler, Schramm and Werner \cite{LSW04} is that this random curve converges to $\rm{SLE}_{8/3}$ from $a$ to $b$ in the domain $\Omega.$ These two conjectures must be considered the principal open problems in the field. If they could be proved, the existence and value of the critical exponents, as predicted by conformal field theory for two-dimensional walks would be proved.

\subsection{Schramm L\"owner Evolution}
For an approachable discussion of SLE$_\kappa,$ see Chapter 15 of \cite{G08}. Here we give a very minimal outline. Let ${\mathbb H}$ denote the upper half-plane. Consider a path $\gamma$ starting at the boundary and finishing at an internal vertex. Then ${\mathbb H}\backslash \gamma$ is the complement of this path, and is a slit upper half-plane. It follows from the Riemann Mapping Theorem that it can be conformally mapped to the upper half-plane. L\"owner \cite{L23} discovered that by specifying the map so that it approaches the identity at infinity, the conformal map so described (actually a family of maps, appropriate to each point on the curve) satisfies a simple differential equation, called the L\"owner equation. The mapping can alternatively be defined by a real function. This observation led Schramm to apply the L\"owner equation to a conformally invariant measure for planar curves. That is to say, the L\"owner equation generates a set of conformal maps, driven by a continuous real-valued function. Scramm's profound insight was to use Brownian motion $B_t$ as the driving function\footnote{It is the only process compatible with both conformal invariance and the so-called domain Markov property.}. So let $B_t,$ $t \ge 0$ be standard Brownian motion on ${\mathbb R}$ and let $\kappa$ be a real parameter. Then SLE$_\kappa$ is the family of conformal maps ${g_t:t \ge 0}$ defined by the L\"owner equation
$$ \frac{\partial}{\partial t} g_t(z) = \frac {2}{g_t(z)-\sqrt{\kappa}B_t}, \,\,\, g_0(z)=z.$$ This is actually called {\em chordal} SLE$_\kappa$ as it describes paths growing from the boundary and ending on the boundary. If $\kappa \le 4$ then the path is almost surely a simple curve, in the upper half plane. Larger values of $\kappa$ lead to more complicated behaviour.

\section{Conclusion}

 Hopefully this rather vague description will convey the flavour of this exciting and powerful development in studying not just two-dimensional SAWs, but a variety of other processes, such as percolation, the random cluster model, and the Ising model. We refer the reader to \cite{B10} for greater detail of both SLE and these applications. Despite these remarkable advances, we still have no real idea how to obtain comparable results for the 3-dimensional model\footnote{This is also true of other classical models, such as the Ising model, the Potts model and percolation.}.

In this article I have only scratched the surface of this topic. More details on the mathematical aspects can be found in \cite{madras-slade} and the recent reviews \cite{BDGS10, B10}. More information on numerical aspects and some applications, particularly to the SAP subset can be found in the monograph \cite{G08}. Another approach to this problem that has not been discussed is to simplify the problem so that it can be solved (see \cite{G08} Chap. 3). Unfortunately most such simplifications involve rendering the model Markovian, which removes a significant feature.

As can be seen, there are many challenging open problems in the field that might capture the imagination of mathematical physicists, computer scientists and probabilists. Perhaps the major open problem is to prove that the scaling limit of SAWs is described by $SLE_{8/3}.$ Even proving the existence of critical exponents for two- or three-dimensional SAWs would be a significant advance.

\section{Acknowledgements}
I would like to thank Nathan Clisby for his valuable comments on the manuscript, and my co-authors of the works described here. Financial support from the Australian Research Council through grant DP120100939 is gratefully acknowledged.


\begin{thebibliography}{10}
\bibitem{DC-H} H Duminil-Copin and A Hammond, Self-avoiding walk is sub-ballistic, arXiv1205:0401v1.
\end{thebibliography}
\end{document}